# Room Temperature Gate-Tunable Non-Reciprocal Charge Transport in Lattice-Matched InSb/CdTe Heterostructures

Lun Li[1#], Yuyang Wu[2#], Xiaoyang Liu[1,3,4#], Hanzhi Ruan[1], Zhenghang Zhi[1,3,4], Jiuming Liu[1], Yong Zhang[1,3,4], Puyang Huang[1], Yuchen Ji[5], Chenjia Tang[5,6], Yumeng Yang[1], Renchao Che[2*], and Xufeng Kou[1,6*]

[1]School of Information Science and Technology, ShanghaiTech University, Shanghai, 201210, China

[2]Department of Materials Science, Fudan University, Shanghai 200438, P. R. China

[3]Shanghai Institute of Microsystem and Information Technology, Chinese Academy of Sciences, Shanghai 200050, China

[4]University of Chinese Academy of Science, Beijing 101408, China

[5]School of Physical Science and Technology, ShanghaiTech University, Shanghai, 201210, China

[6]ShanghaiTech Laboratory for Topological Physics, ShanghaiTech University, Shanghai 201210, China

* Corresponding author. Email: che@fudan.edu.cn and kouxf@shanghaitech.edu.cn

**The manipulation of symmetry provides an effective way to tailor the physical orders in solid-state systems. With the breaking of both the inversion and time-reversal symmetries, non-reciprocal magneto-transport may emerge in assorted non-magnetic systems to enrich spintronic physics. Here, we report the observation of the uni-directional magneto-resistance (UMR) in the lattice-matched InSb/CdTe film up to room temperature. Benefiting from the strong built-in electric field of 0.13 V·nm$^{-1}$ in the hetero-junction region, the resulting Rashba-type spin-orbit coupling and quantum confinement warrant stable angular-dependent second-order charge current with the non-reciprocal coefficient 1-2 orders of magnitude larger than most non-centrosymmetric materials at 298 K. More importantly, this heterostructure configuration enables highly-efficient gate tuning of the rectification response in which the enhancement of the UMR amplitude by 40% is realized. Our results advocate the narrow-gap**

**semiconductor-based hybrid system with the robust two-dimensional interfacial spin texture as a suitable platform for the pursuit of controllable chiral spin-orbit devices and applications.**

Symmetry, which lies at the heart of the universal laws, serves as a fundamental degree of freedom to define the intrinsic properties of a material[1]. Together with spin-orbit coupling (SOC), the breaking of inversion symmetry modifies the Hamiltonian of the charge carriers with anti-symmetric momentum operators, thus leading to non-reciprocal transport phenomena featured by unidirectional *I-V* characteristics[2]. In order to realize such rectification effect, it requires the magneto-chiral anisotropy that originates from either the non-centrosymmetric band structures or hetero-interfaces[3-5]. In this context, the inherent bulk/interfacial Rashba-type SOC would lift the spin degeneracy and trigger the current-induced spin polarization[6-11]. By further applying an external magnetic field to break the time-reversal symmetry, the regulation of the asymmetric spin splitting can be represented by the non-reciprocal electrical response where the longitudinal resistance exhibits unique current direction-dependent signature and its second-order harmonic component scales linearly with both the applied current and the magnetic field[2,4,5,12-23]. Governed by the spin-momentum locking mechanism, this unidirectional magnetoresistance (UMR) can be established without invoking additional magnetic order, hence unfolding a reliable approach to explore the spin-orbit dynamics in a variety of low crystalline symmetry systems including topological insulators/semimetals (*e.g.*, $Bi_2Se_3$[19], $WTe_2$[20]), polar semiconductors (*e.g.*, $BiTeBr$[5], $\alpha$-$GeTe$[21]) and heterostructures (*e.g.*, $LaAlO_3/SrTiO_3$[4,22], Ge/Si[23]).

Considering the amplitude of the Rashba-type UMR signal is closely linked to the spin splitting energy $\Delta\varepsilon \sim \alpha_R \cdot (\Delta \boldsymbol{k} \times \hat{n}) \cdot \boldsymbol{\sigma}$ (where $\alpha_R$ is the Rashba coefficient, $\boldsymbol{k}$ stands for momentum vectors, $\boldsymbol{\sigma}$ is the Pauli operator, and the mirror symmetry is broken along the $\hat{n}$ direction), the spin relaxation time $\tau$, and the Fermi energy $\varepsilon_F$, the key to facilitate the aforementioned non-reciprocal charge transport relies on the optimization of the internal polarization/electrical field as well as the tunability of the Fermi level[4,5,21,22]. Accordingly, with appropriate design of heterostructures, the band bending at the interface can introduce the built-in electric field that ensures strong Rashba-type spin-orbit interaction, and the well-developed quantum well also helps to confine the electron conduction within the two-dimensional channel (*i.e.,* which secures a long phase-coherent/scattering length). In addition, such band engineering allows for the direct modulation of the Rashba SOC strength and the Fermi level through the gate bias, which may cultivate versatile voltage-controlled non-reciprocal functionalities.

Inspired by the above design rules, in this work, we demonstrate the use of the InSb/CdTe heterostructures to achieve the electric field control of non-reciprocal charge transport up to room temperature. As illustrated in Fig. 1a, the integration of the narrow-bandgap InSb thin film with the wide-bandgap CdTe buffer layer takes full advantages of the type-I hetero-junction with a large conduction band offset of $\Delta E_C = 0.31$ eV to generate strong built-in electric field and Rashba coefficient, both of whose values hold the record among reported III-V semiconductor-based heterostructures. Moreover, the nearly identical lattice constants between the two adjacent layers guarantee the formation of an atomically-sharp interface so that the spin-polarized electrons can move through with high carrier mobility. As a result, the giant Rashba field-induced spin splitting energy installed in the InSb/CdTe system gives rise to the enhanced unidirectional magnetoresistance and a gate-tunable non-reciprocal coefficient $\gamma$ in the entire temperature range of 1.5 K $< T <$ 298 K. Our observations showcase the importance of the heterostructure engineering on preserving the non-reciprocal responses at room temperature, and offer feasible strategies for energy-efficient spin-state manipulation.

**Results and Discussion**

**Giant Built-In Electric Field in InSb/CdTe Heterostructures.** Experimentally, high-quality InSb/CdTe thin films were grown on the 3-inch semi-insulating GaAs (111)B wafer by molecular beam epitaxy (MBE) [24], and detailed growth information and sample characterizations are discussed in Supplementary Fig. S1. Figures 1b reveals the cross-sectional high-angle annular dark field (HAADF) profile of the as-grown sample which confirms the absence of macroscopic threading dislocations in the examined region. Meanwhile, the structural characteristics near the InSb/CdTe junction are visualized by the atomic-resolution X-ray energy dispersive spectroscopy (EDS) mapping in Fig. 1c. It is clearly seen that all the [In, Sb, Cd, Te] atoms rigorously follow the zinc-blende configuration along the [111] direction, and the defect-free hetero-interface is well-defined with negligible inter-mixing among the constituent elements, hence validating the epitaxial growth of the single-crystalline heterostructures. Furthermore, the built-in electric field due to the interfacial band bending is quantified through the off-axis electron holography. Owing to the non-magnetic property of the InSb/CdTe sample, the phase shift of the incident electron beam is only affected by the electric field signal inside the sample. Therefore, the distribution of the internal electric potential is attained from the relevant phase diagram (see Supplementary Fig. S2)[25]. As emphsized in Fig. 1d, the peak position of the electric field is indeed found to locate at the InSb/CdTe interface, and its corresponding maximum intensity ($E_{max} = 0.13$ V·nm$^{-1}$ at $T = 298$ K) is much higher than both the conventional GaAs/Al$_x$Ga$_{1-x}$As quantum wells (0.015 ~ 0.045 V·nm$^{-1}$) and the InAs/GaSb superlattice (0.08 V·nm$^{-1}$) [26]. In the meantime, the full width at half maximum (FWHM) of the electric-field spectrum is about 8 nm, and the electric potential energy difference of 0.3 eV is obtained by integrating the electric field within the hetero-junction region. These results are in a good agreement with the band diagram simulated by technology computer-aided design (TCAD)[27], again validating the existence of the giant Rashba effect in the lattice-matched InSb/CdTe system.

**Room-Temperature Rashba-Type Unidirectional Magnetoresistance.** From our previous study, we have detected distinct spin-torque ferromagnetic resonance (ST-FMR) signals from the hybrid Py-InSb/CdTe films at $T = 298$ K[28], which suggest productive spin polarization in the InSb/CdTe channel up to room temperature. To further investigate the inversion asymmetry-related effects, we performed angular-dependent magneto-transport measurements on the μm-size six-probe Hall-bar device patterned from the MBE-grown InSb(30

nm)/CdTe(400 nm) sample (see Methods). Given that the interfacial Rashba SOC-induced pseudo-magnetic field $\boldsymbol{B}_{\text{eff}}$ is tightly locked to the product of the momentum operator and the built-in electric field (*i.e.*, $\boldsymbol{B}_{\text{eff}} = \boldsymbol{\alpha}_{\text{R}} \cdot (\boldsymbol{k}_x \times \hat{\boldsymbol{z}})$), it would influence the in-plane spin-dependent scattering. In this regard, by applying the low-frequency AC current $I = I_0 \cdot \sin(\omega t)$ along the +*x* direction while rotating the external magnetic field in the *xy*-plane (*i.e.*, the angle between $\boldsymbol{B}$ and $I$ is defined as $\phi$), the 2$^{\text{nd}}$-harmonic components of both the longitudinal ($V_{xx}^{2\omega}$) and vertical ($V_{xy}^{2\omega}$) voltages were recorded using the standard lock-in technique. Figure 2a presents one set of the non-linear resistance pairs ($R_{xx}^{2\omega} = V_{xx}^{2\omega}/I_0$, $R_{xy}^{2\omega} = V_{xy}^{2\omega}/I_0$) under a given input condition of $B = 14$ T and $I_0 = 400$ μA, where two sinusoidal slopes with the same period of 360° but a phase difference of 90° are observed in the $R^{2\omega} - \phi$ diagram at $T = 298$ K. By excluding the thermal contribution from the Nernst effect (*i.e.*, the current-induced vertical temperature gradient also causes the non-linear transport response, see Supplementary Fig. S3), the Rashba-type UMR behavior is subsequently demonstrated in Figs. 2b-d, namely the extracted 2$^{\text{nd}}$-harmonic magnetoresistance $R_{\text{UMR}}^{2\omega}$ reaches its highest (lowest) state at $\phi = 90°$ (270°) whereas it vanishes when $\phi = 0°$ and 360°. Similar with other non-centrosymmetric materials, the UMR amplitude $\Delta R_{\text{UMR}}^{2\omega}$ in our InSb/CdTe sample also displays linear relations with the magnetic field (up to 14 T) and the charge current (up to 400 μA), therefore manifesting the $\Delta R_{\text{UMR}}^{2\omega} \propto B \cdot I_0$ nature of the non-reciprocal transport[4,5,19,20,21,23].

To understand the interplay between the Rashba-type spin splitting and $\Delta R_{\text{UMR}}^{2\omega}(\varphi, B)$, it is firstly recalled that the Fermi level position within the InSb/CdTe junction region resides above the conduction band $\Gamma_6$ of InSb (see Supplementary Fig. S4) [27,29]. By applying the current along the +*x*-axis, the interfacial conducting electrons are spin-polarized toward the +*y* direction, and the 2D Rashba Hamiltonian under an in-plane magnetic field is thus given by:

$$H = \frac{\hbar(k_x^2 + k_y^2)}{2m^*} + (\alpha_{\text{R}} k_x) \cdot \sigma_y - (\frac{1}{2} g\mu_B B \sin\phi) \cdot \sigma_y$$

where $\hbar$ is the reduced Planck constant, $m^* = 0.03\, m_e$ is the electron effective mass[30], $g = -42$ is the in-plane g-factor[31,32,33], $\mu_B$ is the Bohr magneton, and the Rashba coefficient $\alpha_R = 0.4$ eV·Å is deduced from the weak anti-localization data[17]. In reference to such database, a semi-empirical energy band dispersion with $|\alpha_R k_F| = 9$ meV and $\Delta\varepsilon_z = g\mu_B B = -34$ meV is shown in the top panel of Fig. 2e. In accordance with the non-linear second-order spin-orbit coupled magneto-transport model, the $\Delta k$-induced spin sub-band displacement tilts the electron distribution at the Fermi surface, and the in-plane magnetic field disequilibrates the spin-up and spin-down densities in the system[19,21,22,23]. Accordingly, the second-order variation of the electron distribution $\delta f_2$ results in a non-zero second-order charge current $J'_c \propto E_x^2 \cdot B\sin\phi \propto \Delta R^{2\omega}_{UMR}$[5,19,21]. As exemplified in the bottom panel of Fig. 2e, the ($+B_y$, $+E_x$) condition initiates a majority of electrons with a $+\sigma_y$ spin state, which in turns drives the induced $J'_c(+B_y, +E_x^2)$ along the $-x$ direction; on the contrary, the reversal of the applied magnetic field to $-B_y$ switches the sign of $J'_c(-B_y, +E_x^2)$ toward the $+x$ direction. Consequently, a high (low) resistance state is expected when the magnetic field is parallel (anti-parallel) to the spin polarization direction, consistent with the UMR results of Fig. 2b. Additionally, it is noted that $\Delta R^{2\omega}_{UMR}$ evolves non-monotonously from $T = 1.5$ K to 298 K (inset of Fig. 2b), which may possibly correlate with the temperature-dependent density of states at the Fermi surface[21], yet quantitative electronic band structure calculations of InSb with the presence of the Rashba splitting under external fields are needed for future study.

Based on the same principle discussed earlier, we are able to map the spin texture of the InSb/CdTe heterostructures by tracing the non-reciprocal transport signal with respect to the current and magnetic field directions. For instance, Figs. 3a-c list three UMR slopes measured by rotating the $B_{yz} = 14$ T in the $yz$-plane while the angles between input current and the $x$-axis are chosen as $\phi = 90°$, $60°$, and $30°$, respectively. All $R^{2\omega}_{UMR}$ curves exhibit sinusoidal angular dependence with a period of 360°, and the normalized peak amplitude ratio of $\Delta R^{2\omega}_{UMR}(\phi)/\Delta R^{2\omega}_{UMR}(\phi = 90°)$ follows the $\sin\phi$ relation (Fig. 3d). Comparing with the in-plane reference data re-captured from Fig. 2b (red line), we may conclude that the non-reciprocal response in our sample mainly stems from the two-dimensional conduction channel at the hetero-interface where the spin-momentum locking nature of the Rashba effect only produces spin polarization strictly orthogonal to $\boldsymbol{k} \times \boldsymbol{E_z}$ (*i.e.,* counter-clockwise spin chirality of the outer branch in the *k*-space), and there is no out-of-plane spin component since the $R_{\text{UMR}}^{2\omega}$ contours in the *xy* and *yz*-planes are identical to each other (Fig. 3a). Besides, we have also carried out the angular-dependent second-order harmonic measurements on a series of Hall-bar devices fabricated along different crystal orientations, and the same obtained UMR results disclose the isotropic feature of the energy band (see Supplementary Fig. S5).

In order to evaluate the strength of the non-reciprocal charge transport, here we choose the non-reciprocal coefficient $\gamma = 2\Delta R_{\text{xx}}^{2\omega}/(B \cdot I_0 \cdot R_0)$ as the main figure-of-merit. Strikingly, from the summarized temperature-dependent $\gamma$ data in various non-centrosymmetric systems (Fig. 3e), our InSb/CdTe thin film obviously distinguishes itself from this benchmark chart in terms of both a large $\gamma \sim 0.52\ A^{-1}T^{-1}$ and the robust thermal stability. In particular, guided by the criteria of $\gamma \propto g\alpha_R \tau^2/\varepsilon_F$, it is evident that even though the polar semi-metals and semiconductors (*e.g.,* WTe$_2$[20], α-GeTe[21], and Bi-helix[34]) inherit giant bulk Rashba coefficients as high as 2~5 eV·Å, yet the metallic electrical properties (*i.e.,* high carrier density and bulk scattering) inevitably limit the magnitude of $\gamma$ below $10^{-2}\ A^{-1}T^{-1}$ at room temperature. On the other hand, although semiconductor and oxide heterostructures (*e.g.,* SiO$_2$/Si[35], Ge/Si[23], and LaAlO$_3$/SrTiO$_3$[4]) host the two-dimensional electron gas channels at low temperatures, the weak band bending and small spin splitting energy (3~5 meV) make the systems vulnerable to thermal fluctuation, that is the UMR signals attenuate dramatically with increased temperatures and are barely observable at room temperature. As a result, only the combinations of strong interfacial Rashba SOC, high electron mobility ($\mu_e$ = 3000 cm$^2$V$^{-1}$s$^{-1}$), and relatively small Fermi energy ($\varepsilon_F$ = 58 meV) have endowed the lattice-matched InSb/CdTe system an optimized $\gamma$-value over the entire temperature range.

**Electric Field Control of the Non-Reciprocal Transport.** Thanks to the effective gate-tuning of the Rashba SOC strength in the InSb/CdTe heterostructures[27], we further deposited a 50 nm $Al_2O_3$ dielectric layer on top of the InSb(30 nm)/CdTe(400 nm) wafer and designed the top-gated Hall-bar structure using nano-fabrication process (Fig. 4a). The gate-dependent UMR results of the fabricated device at $T = 1.5$ K are presented in Fig. 4b. By changing the gate voltage ($V_g$) from -4 V to +6 V, the normalized $R_{norm}^{2\omega} = \frac{R_{UMR}^{2\omega}(V_g)}{\Delta R_{UMR}^{2\omega}(V_g=0)}$ curves all retain the original $\sin\phi$ line-shape while the non-reciprocal amplitude $\Delta R_{UMR}^{2\omega}(V_g)$ successively shrinks from 0.67 Ω to 0.47 Ω (i.e., 40% gate-voltage modulation efficiency), and the same negatively-correlated $\gamma_{norm} - V_g$ trend can be observed from 1.5 K up to 298 K, as shown in Fig. 4c.

According to the TCAD-simulated spatial distribution profile of the electron density across the $Al_2O_3$/InSb/CdTe structure (Fig. 4d), it is seen that the top-gate configuration enables a highly-effective charge carrier regulation within the InSb layer in view of its semiconducting nature. Specifically, the applied negative bias introduces a large number of holes (*i.e.*, $Q = C_{ox} \cdot V_g$, where $C_{ox}$ is the capacitance of the $Al_2O_3$ layer) into the *n*-type InSb layer, thereafter driving the Fermi level position toward the InSb band gap center. Upon the occurrence of the bulk channel depletion, a majority of the charge current would conduct through the underlying InSb/CdTe interfacial channel with a strong built-in potential. Under such circumstances, both the enhanced Rasha SOC strength[27] and lowered Fermi energy help to amplify the non-reciprocal transport response in the negative bias region. In contrast, when $V_g > 0$ V is applied, the induced excessive electrons bring about the bulk-dominated conduction scenario with a damped $\gamma$-value (*i.e.,* even an electron accumulation layer may be formed near the $Al_2O_3$/InSb interface, the itinerant carriers would experience the suppressed $\alpha_R$ and $\tau$ because of the less ideal interface quality). Besides, we need to point out that the increase of temperature promotes the thermal activation with a more dominant bulk conduction in the gated InSb/CdTe device, hence making the gate-voltage modulation less effective at higher temperatures (Fig. 4c). In addition to the UMR effect arisen from the second-order charge current, the first-order anisotropic magneto-resistance (AMR) of our InSb/CdTe heterostructures also unveils a unique gate-dependent trait. As identified in Fig. 4e, the AMR amplitude of the Al$_2$O$_3$(50 nm)/InSb(30 nm)/CdTe(400 nm) device gets boosted significally with the increase of the applied gate voltage (*i.e.,* corresponding to the transition from the two-dimensional hetero-junction conduction to the bulk InSb transport). Concurrently, more pronounced AMR signals are also discovered in thicker InSb samples (see Supplementary Fig. S6), which implis the origin of this W-shape AMR effect may be associated with the bulk property of the InSb(111) material[36-40].

**Conclusion**

In conclusion, we have investigated the non-reciprocal magneto-transport phenomenon induced by the interfacial Rashaba SOC in the InSb/CdTe heterostructures. The achievements of large Rashba coefficient, long phase-coherent length, and adjustable Fermi energy altogether contribute to the remarkable UMR response whose coefficient $\gamma \sim 0.52\ A^{-1}T^{-1}$ is well-maintained up to room temperature. Furthermore, the systematic angular- and gate-dependent measurements not only manifest the isotropic spin-momentum locking energy band with and in-plane counter-clockwise spin texture, but also excel at salient gate-controlled capability on tuning the UMR (InSb/CdTe hetero-junction transport) and AMR (InSb bulk conduction) components. Our results highlight the indispensable role of band engineering on strengthening the spin-orbit interaction, and the lattice-matched material integration concept may set up a general framework for the design of gate-controllable low-power spin-orbitronics applications.

**Methods**

**Sample Growth and structural characterization**. The InSb/CdTe heterostructures growth was carried out in an ultra-high vacuum DCA dual-chamber MBE system with the base pressure of $3 \times 10^{-11}$ Torr. Semi-insulating ($\rho > 10^6$ Ω·cm) 3-inch GaAs (111) B wafers were pre-annealed in the growth chamber at up to 570 °C in the Te-protected environment to remove the native oxide. Subsequently, a two-step growth procedure was adopted for the CdTe growth with relatively low growth rate of 1.1 Å·s$^{-1}$ to obtain a smooth surface morphology. After a moderate post-annealing at 355 °C, the CdTe film was exchanged onto another substrate holder and later transferred in the III-V chamber. For the subsequent InSb layer growth, the substrate temperature was kept at 300 °C and the growth rate was fixed as 3.5 Å·s$^{-1}$. During the entire epitaxial growth, the beam flux monitor was used to calibrate element flux rate and *in-situ* RHEED was applied to monitor the real-time growth conditions. After sample growth, atomic force microscopy(AFM) were performed to observe surface morphology of as-grown InSb/CdTe sample. In addition, XRD and HRSTEM were applied to check crystal phase and quality.

**Device Fabrication.** The 3-inch InSb/CdTe wafers were cut into small pieces about one centimeter. Then several μm-size six-terminals Hall bar devices along different direction were fabricated on one piece of sample using conventional photolithography and ion-beam etching methods. Ti/Au (50nm/250nm) was deposited by *e*-beam evaporation to form the ohmic contacts after etching. For top-gated FET devices, a 50 nm thick high-κ Al$_2$O$_3$ dielectric layer was deposited by atomic layer deposition (ALD) at 150°C, and the contact metals are made of Ti/Au/Pt (10nm/240nm/50nm).

**Transport Measurement.** The magneto-transport measurements on both the Hall devices and top-gate FET devices were performed with a He-4 refrigerator (Oxford TeslatronPT system). AC current with frequency varied during the measurements were applied on the devices by Keithley source meters, then the first and second harmonic voltage were measured by lock-in amplifiers while several experimental variables such temperature, magnetic field, rotation angle, and current amplitude were varied during the measurements.

**Device Simulation.** The Sentaurus TCAD device simulator from Synopsys was used to investigate the band diagram and carrier density profile of the top-gated InSb/CdTe device. A typical Al$_2$O$_3$(50 nm)/InSb(30 nm)/CdTe(400 nm) MOS structure was defined by the SDE tools. The dielectric constant of Al$_2$O$_3$ was set to 8.6 in reference to the capacitance test result of our ALD grown $Al_2O_3$ thin film. During the simulation process, the gate electrode was placed at the top of the $Al_2O_3$ layer and ground electrodes were placed at the right/left edge of InSb channel. The designed device structure was later exported to the Sdevice tool in which the 2D Poisson solver included both electron and hole was performed to obtain the real-space band diagram and carrier distribution under different gate voltage bias.

**Acknowledgements**

This work is sponsored by the National Key R&D Program of China under the contract number 2017YFB0305400, National Natural Science Foundation of China (Grant No. 61874172 and 92164104), and the Major Project of Shanghai Municipal Science and Technology (Grant No. 2018SHZDZX02). X.F.K acknowledges the support from the Merck POC program and the Shanghai Rising-Star program (Grant No. 21QA1406000). Y.M.Y acknowledges the support from Shanghai Pujiang Program (Grant No. 20PJ1411500).

**Author contributions**

X. F. Kou and R.C Chao conceived and supervised the study. H.Z Ruan, Y.C Ji, and C.J Tang grew the samples. L. Li, X.Y Liu, J.M Liu, and P.Y Huang performed device fabricaiton and conducted the transport measurements. Y.Y Wu carried out the microscopy characterizations. L. Li, X.Y Liu, and Y.M Yang analyzed the transport data. Z.H Zhi and Y. Zhang conducted the TCAD simulations. L. Li, X.Y Liu and X. F. Kou wrote the manuscript. All authors discussed the results and commented on the manuscript.

**Competing financial interests**

The authors declare no competing financial interests.

**Figures**

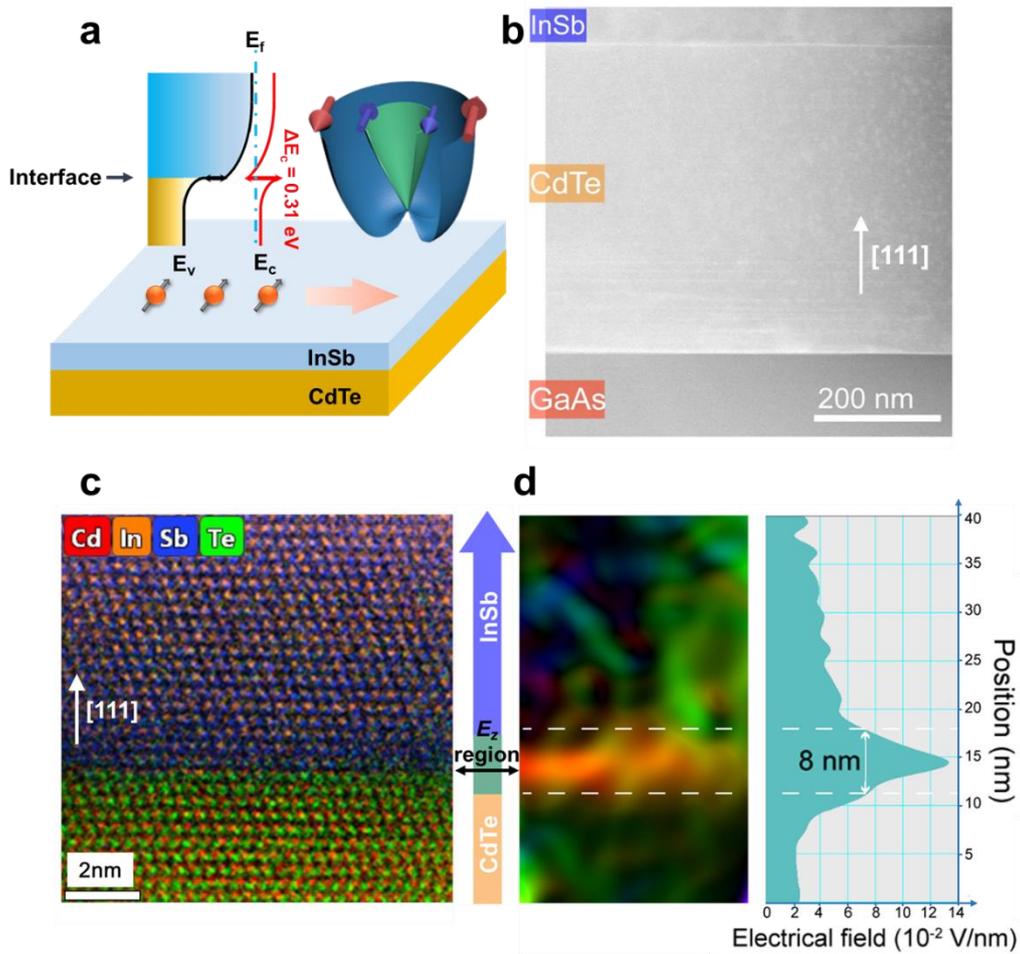

**Figure 1 | Structural characterizations of the InSb(30 nm)/CdTe(400 nm) heterostructures. a**. Schematic of the InSb/CdTe heterostructure and relevant energy band diagram. The well-established band bending at the InSb/CdTe heter-junction causes strong Rashba-type spin-orbit coupling on the spin sub-bands and forces the interfacial electrons to spin-polarize perpendicular to the current direction. **b**. cross-sectional HAADF image displays the intact interface of the InSb/CdTe sample. **c**. Atomic-resolution EDS mapping near the InSb/CdTe interface region. **d**. Electron holography profile confirms the exisistence of the strong built-in electric field with the peak intensity of 0.13 V·nm$^{-1}$ and the average width of 8 nm.

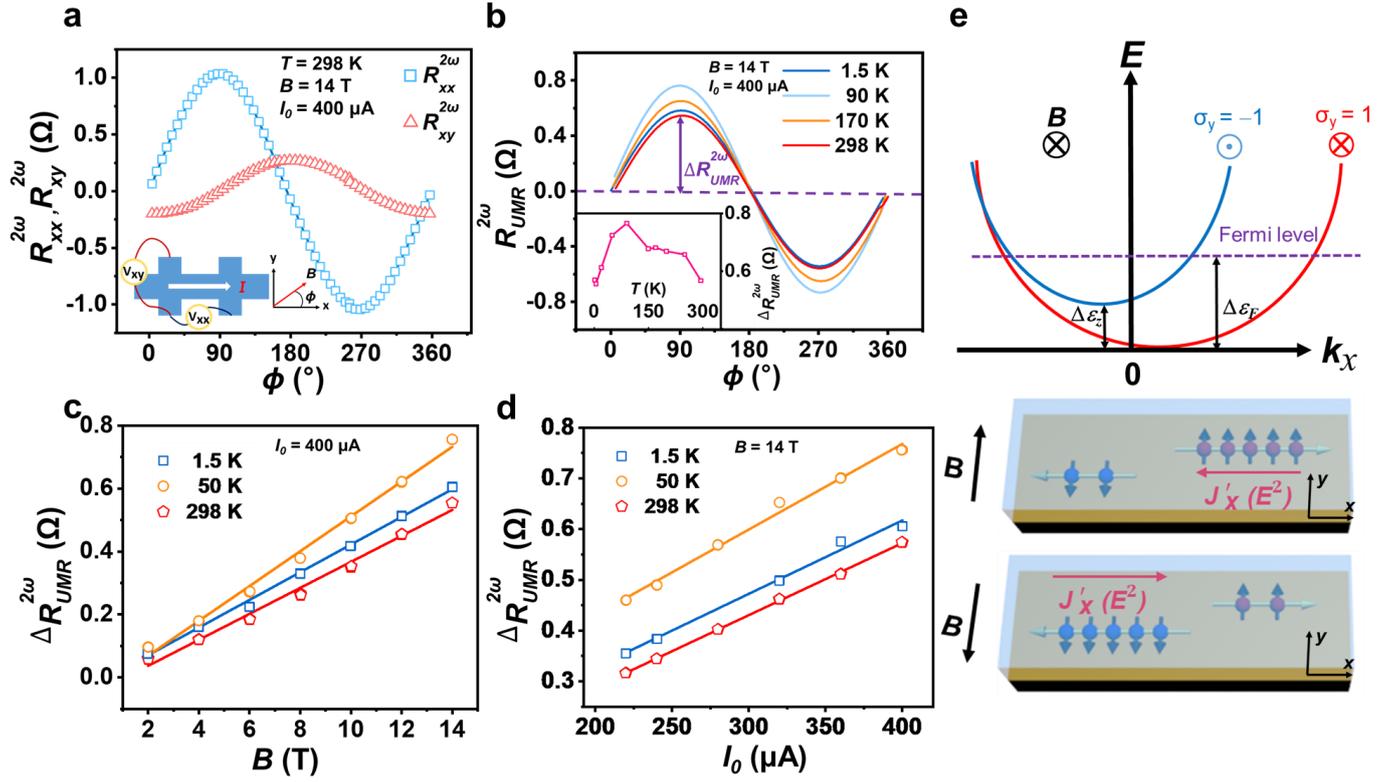

**Figure 2 | Rashba-type unidirectional magneto-resistance effect in the InSb/CdTe heterostructures. a.** Angular-dependent 2nd-harmonic longitudinal ($R_{xx}^{2\omega}$) and vertical ($R_{xy}^{2\omega}$) resistances of the InSb(30 nm)/CdTe(400 nm) sample at $T$ = 298 K. Both data were recorded using standard lock-in technique while the applied magnetic field continuously rotates in the *xy*-plane. The inset image illustrates the measurement setup and the rotation angle definition. **b.** The Rashba spin splitting-induced sinusoidal UMR component $R_{UMR}^{2\omega}$ probed from 1.5 K to 298 K. Inset: the UMR amplitude $\Delta R_{UMR}^{2\omega}$ displays an non-monotonical dependence on temperature. **c.** The UMR amplitude is proportional to the applied in-plane magnetic field under a fixed current amplitude of $I_0$ = 400 μA. **d.** Linear current-dependent $\Delta R_{UMR}^{2\omega}$ curves under the same $B$ = 14 T. **e.** Band structure of the InSb conduction band with the magnetic field along the +*y*-direction. Owning to the negative *g*-factor of InSb, the negative Zeeman energy $\Delta\varepsilon_z = -34$ meV drives the spin-up sub-band (red) below the spin-up sub-band (blue). The Fermi level $\Delta\varepsilon_F = 58$ meV locates above the conduction band minimum (CBM), and the cross point of the sub-bands $\Delta\varepsilon_{cp} = 209$ meV is much higher than CBM. Therefore, the second-order current is enhanced (suppressed) by the in-plane magnetic field parallel (antiparallel) to the spin

polarization.

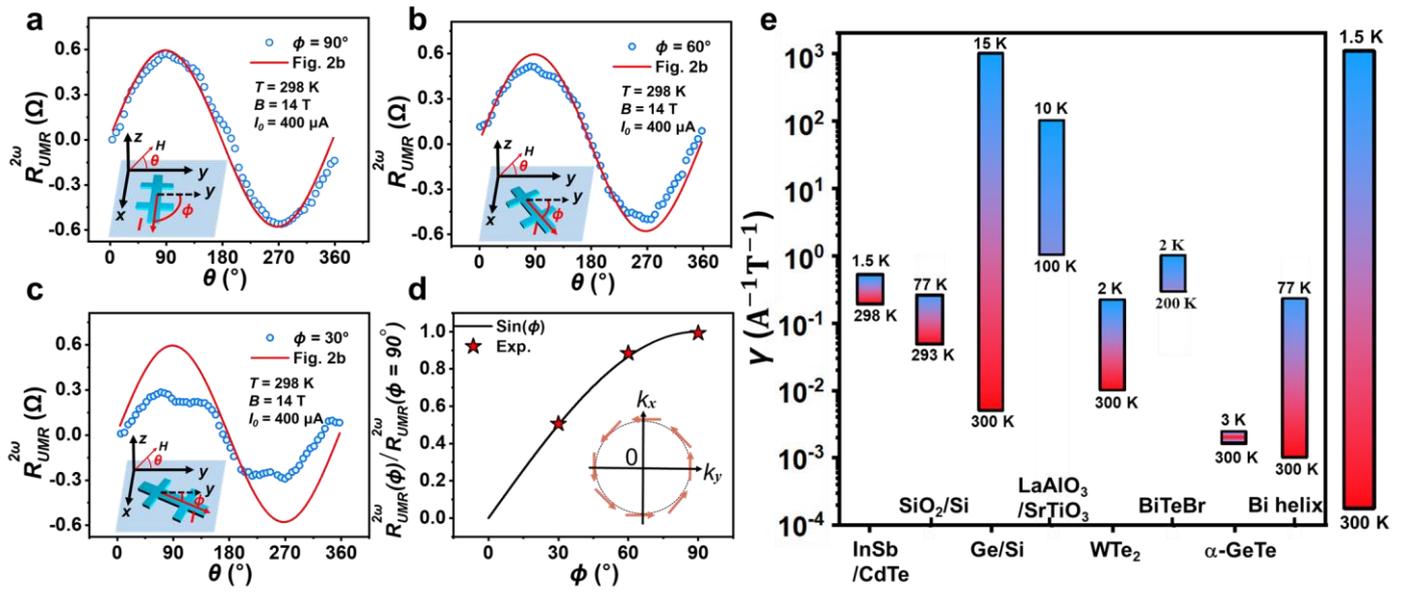

**Figure 3 | Mapping non-reciprocal transport responses in the InSb/CdTe heterostructures. a-c**. angular-dependent UMR results of the InSb(30 nm)/CdTe(400 nm) sample at $T$ = 298 K. The applied magnetic field rotates within the *yz*-plane while the angles between input current and the *x*-axis are chosen as $\phi$ = 90°, 60°, and 30°, as shown in the inset schematic diagrams. **d**. The normalized UMR amplitude of devices is depicted by the rigorous $\sin\phi$ relation, indicating the absence of any out-of-plane spin polarization component. Inset: Counter-clockwise spin chirality of the interfacial electrons in the *k*-space. **e**. Comparison of the non-reciprocal coefficient in various non-centrosymmetric material systems. Our InSb/CdTe sample stands out from the category by exhibiting a thermal stable non-reciprocal coefficient with a large $\gamma$-value at room temperature.

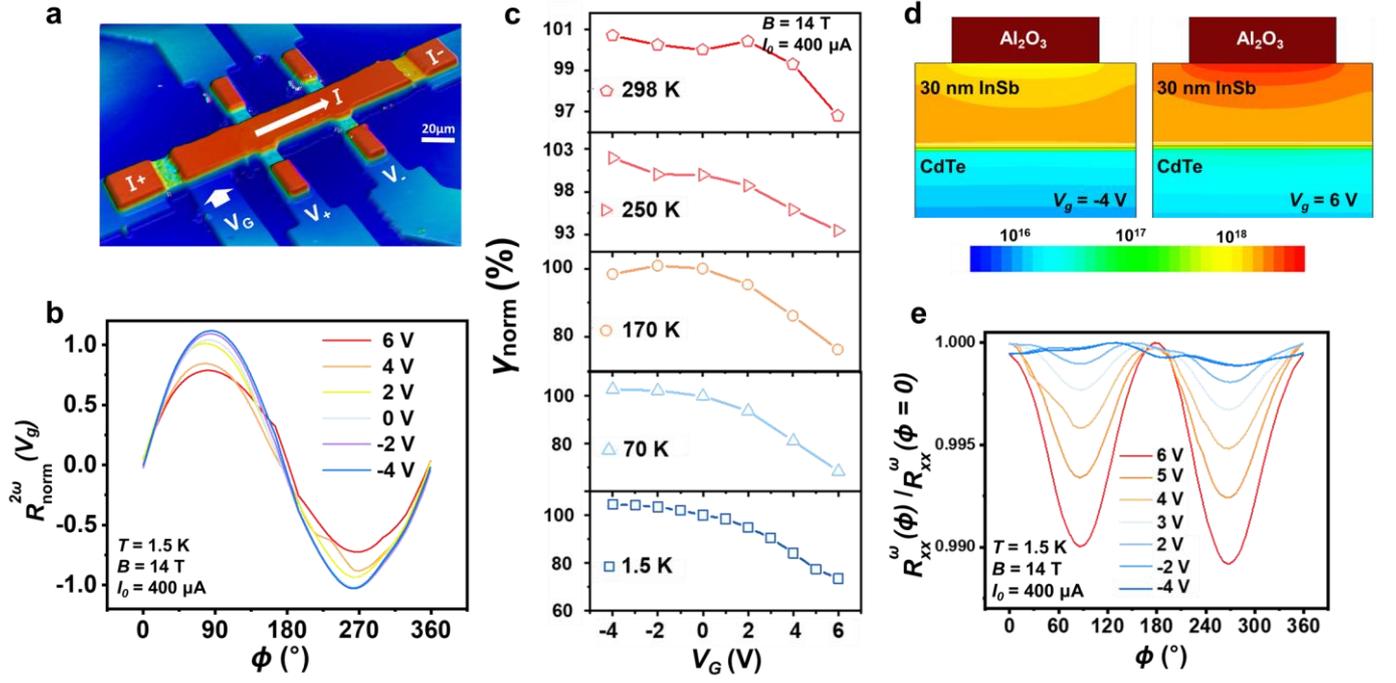

**Figure 4 | Electric field control of the 1st/2nd-order magneto-resistance strength in the top-gated InSb(30 nm)/CdTe(400 nm) device. a**. Three-dimensional confocal laser scanning image of the fabricated device with the gate size of 100 μm (length) × 20 μm (width). **b**. Gate-controlled normalized UMR results of the device by applying the magnetic field within the *xy*-plane at room temperature. **c**. Electric field control of the normalized $\gamma_{\text{norm}} = \gamma(V_g)/\gamma(V_g = 0\text{ V})$ at different temperatures varying from 1.5 K to 298 K. All $\gamma_{\text{norm}}$ curves follow the same negatively-correlated evolution trend with respect to the applied voltage, and the weakend gate tunablity implies the increase of InSb bulk conduction at elevated temperatures. **d**. TCAD-simulated carrier density distribution profile in the Al$_2$O$_3$/InSb/CdTe tri-layer structure at $V_g = -4$ V (left panel) and 6 V (right panel). **e**. Gate-dependent first-order anisotropic magneto-resistance of the same device. Unlike the $R^{2\omega}_{\text{norm}} - V_g$ behavior, the AMR amplitude gradually enlarges as $V_g$ changes from −4 V to 6 V, reflecting a different mechanism related to the InSb bulk property.